\documentclass[pra,twocolumn,amsmath,amssymb,superscriptaddress,showpacs]{revtex4-1}

\usepackage[caption=false]{subfig}
\usepackage{graphicx}
\usepackage{mathrsfs}
\usepackage[colorlinks]{hyperref}
\usepackage{braket}

\newcommand{\ms}[1]{\mbox{\scriptsize #1}}

\begin{document}

\title{Microwave-to-optical frequency conversion using a cesium atom coupled to a superconducting resonator}

\author{Bryan T. Gard}
\email{bryantgard1@gmail.com}
\affiliation{U.S. Army Research Laboratory, Computational and Information Sciences Directorate, Adelphi, Maryland 20783, USA}
\affiliation{Hearne Institute for Theoretical Physics, Louisiana State University, Baton Rouge, LA 70803, USA}
\author{Kurt Jacobs}
\affiliation{U.S. Army Research Laboratory, Computational and Information Sciences Directorate, Adelphi, Maryland 20783, USA}
\affiliation{Hearne Institute for Theoretical Physics, Louisiana State University, Baton Rouge, LA 70803, USA}
\affiliation{Department of Physics, University of Massachusetts at Boston, Boston, MA 02125, USA}
\author{R. McDermott}
\affiliation{Department of Physics, University of Wisconsin-Madison, 1150 University Avenue, Madison, Wisconsin 53706}
\author{M. Saffman}
\affiliation{Department of Physics, University of Wisconsin-Madison, 1150 University Avenue, Madison, Wisconsin 53706}

\begin{abstract}
A candidate for converting quantum information from microwave to optical frequencies is the use of a single atom that interacts with a superconducting microwave resonator on one hand and an optical cavity on the other. The large electric dipole moments and microwave transition frequencies possessed by Rydberg states allow them to couple strongly to superconducting devices. Lasers can then be used to connect a Rydberg transition to an optical transition to realize the conversion. Since the fundamental source of noise in this process is spontaneous emission from the atomic levels, the resulting control problem involves choosing the pulse shapes of the driving lasers so as to maximize the transfer rate while minimizing this loss. Here we consider the concrete example of a cesium atom, along with two specific choices for the levels to be used in the conversion cycle. Under the assumption that spontaneous emission is the only significant source of errors, we use numerical optimization to determine the likely rates for reliable quantum communication that could be achieved with this device. These rates are on the order of a few Mega-qubits per second. 
\end{abstract} 


\maketitle 

\section{Introduction} 

Superconducting circuits are a promising candidate for realizing quantum information processing~\cite{devo13,nori05}. These circuits operate at microwave frequencies while long-distance communication is performed with light in the optical band. Enabling superconducting quantum information processors to communicate on a large-scale quantum network will therefore require the quantum-coherent conversion of microwaves to optical frequencies and vice versa~\cite{salomon1999, Hollberg2005, Hollberg2005b, schnatz2009, Tian10, Taylor11, Tian12, taylor12, morigi15}. Converting quantum information contained in photons with one frequency to photons with another also has potential applications to efficient atom-photon coupling~\cite{atomphoton}, fast quantum gates~\cite{quantumgates, quantumgates2}, measurement schemes~\cite{statedet, statedet2}, astronomy~\cite{astro1}, frequency standards~\cite{freq1,astro1}, and quantum computing~\cite{qcomp1,qcomp2,qcomp3,kimble08}. 

There are presently three prominent proposals for enabling microwave-to-optical frequency conversion in the fully coherent quantum regime. In these approaches the conversion is mediated by, respectively, a nano-mechanical resonator~\cite{Tian10, Taylor11, Tian12, Barzanjeh12, Clerk12, Bochmann13, Bagci14, Andrews14, Lecocq16}, an ensemble of trapped atoms~\cite{Verdu09, taylor12, morigi15, jaksch16, jaksch17}, and an ensemble of spins in a solid (e.g., NV-centers in diamond)~\cite{Imamoglu09, Marcos10, Wu10, Kubo10, Amsuss11}. A number of experiments have already demonstrated proof-of-principle conversion between microwaves and optical frequencies using a nano-mechanical system~\cite{Bochmann13, Bagci14, Andrews14, Lecocq16}. We also note that the electro-optical modulators may provide a fourth route to frequency conversion~\cite{Tsang10b, Tsang11b}. 

Analyses of the use of trapped atoms to perform the conversion have so far focussed on the use of an ensemble (a cloud of atoms trapped in an optical lattice). Here we consider the use of a single atom that is trapped in an optical cavity so as to interact strongly with a single mode of the cavity (the configuration often referred to as ``cavity-QED''). We make our analysis concrete by considering the cesium atom, being a primary candidate for frequency conversion, and examine two explicit configurations of cesium levels that could be used to perform the conversion. To explore the potential of this scenario we use numerical optimization to obtain laser pulse shapes that minimize loss while maximizing the conversion speed. We find that using such time-dependent laser pulses greatly enhances the conversion efficiency. Employing a second optimization we calculate from the resulting conversion speed and efficiency the quantum communication rate, being the rate at which quantum information (measured in qubits) can be converted reliably (without error)~\cite{wilde13}. 

\begin{figure}[t]
\leavevmode\includegraphics[width=\columnwidth]{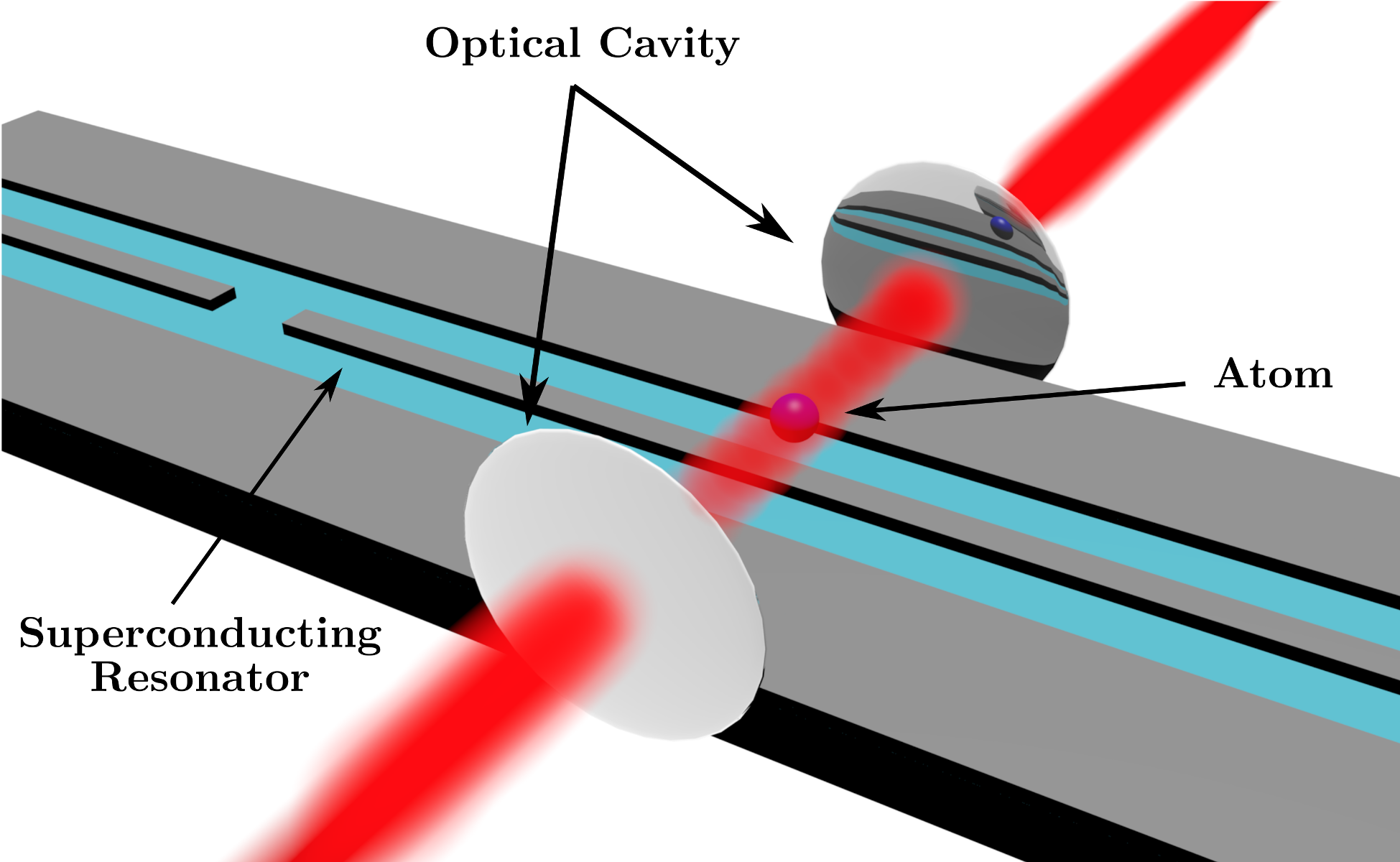}
\caption{The physical layout of the device. A single atom is trapped near a surface containing a superconducting waveguide resonator. The atom is also positioned between two mirrors that form an optical cavity. The role of the atom is to coherently convert photons between the resonator and the optical cavity.} 
\label{system_pic} 
\end{figure}

The use of atoms to convert between microwave photons and optical photons is enabled by the fact that Rydberg transitions --- transitions between highly-excited atomic states --- possess both microwave frequencies and large dipole moments. The latter allows them to interact electrically with superconducting elements with which they are co-located. Thus the proposed device, depicted in Fig.~\ref{system_pic}, consists of a cesium atom in which a Rydberg transition couples to a coplanar waveguide resonator fabricated on a surface and an optical transition couples to an optical cavity. Note that the optical conversion places the optical photon in the telecom band, an ideal wavelength for propagation through an optical fiber. By driving the atomic levels so as to connect the microwave transition to the optical transition, the atom can be used to take a single qubit encoded in the lowest two Fock states of the microwave mode and place it in a mode of the optical cavity. We are interested in the speed (qubits per second) with which quantum information could be reliably transmitted using such a device. It is important to note that the maximum fidelity that can be obtained for the conversion of a single qubit does not limit the fidelity with which quantum information can be transmitted, but rather it limits the rate at which this information can be transmitted reliably. Since the unavoidable source of noise in our conversion process is loss from spontaneous emission, we are interested in the limit that this imposes on the (reliable) transmission rate. To maximize this rate we must choose the laser pulses that mediate the cycle between the Rydberg and optical transitions so as to minimize the loss probability whilst also minimizing the time taken by the cycle. We will find that there is a trade-off between the time taken by the cycle, $\tau$, and the resulting loss probability, $p$, owing to the adiabatic nature of the optimal transfer protocols. 

In Section~\ref{secQTR} we describe how the transmission rate for reliable quantum communication is obtained from the loss probability $p$ and transfer time $\tau$. In Section~\ref{secPS} we describe the physical system, the Hamiltonian and master equation that we use to model it. We also present the two configurations (or ``cycles'') of cesium levels that we use for implementing the transfer. In Section~\ref{secTPO} we elucidate various aspects of the dynamics  that provide insights into the expected behavior and guide the choice of parameters. In particular we discuss the role of the damping rate of the optical cavity that interfaces with the optical communication link and the adiabatic mechanism that can be exploited to minimize loss. In Section~\ref{nopt} we discuss the approach we took to simulating the system and using numerical search methods to find optimal transfer protocols. In Section~\ref{secR} we display the dynamics of an example transfer protocol, and present our results for the minimum loss probabilities and maximum transmission rates achievable with the two cesium configurations. We also present results regarding the robustness of the transfer protocols to fluctuations of the driving lasers, and finish with some concluding remarks. 

\section{Quantum transmission rate}
\label{secQTR}

Because there is a non-trivial relationship between the probability that the excitation is lost during the conversion, $p$, and the resulting quantum communication rate, $\mathcal{R}$, we now give a brief summary of this relationship. The quantum equivalent of Shannon's noisy coding theorem states that the rate at which qubits can be sent with arbitrarily high fidelity is given by multiplying the rate at which the qubits are physically transmitted by a quantity called the \textit{coherent information}, a specific fidelity-like measure of the error for each transmitted qubit~\cite{wilde13}. The coherent information lies between zero and unity, and is the amount of quantum information (in qubits) that can be sent reliably per physical qubit. 

Let us denote the vacuum state of a single (microwave or optical) mode by $|0\rangle$ and the 1-photon Fock state by $|1\rangle$. To convert a state 
\begin{equation}
   |\psi\rangle = \alpha |0\rangle + \beta |1\rangle 
\end{equation}
from microwave to optical frequency involves i) beginning with the microwave mode in state $|\psi\rangle$ and the optical mode in state $|0\rangle$, and ii) transferring the excitation from the microwave to the optical mode. In the scheme we consider here, the physical transfer process involves the microwave cavity exciting the Rydberg transition, at which point the atom is able to complete a loop (cycle) that starts and ends at a ground state, emitting a photon into the optical cavity as part of the cycle. Thus if the initial state of the microwave cavity is $|0\rangle$ no cycle is performed and there is no possibility of spontaneous emission. If the initial state is $|1\rangle$ then a spontaneous emission during the loop will steal the excitation and leave the optical cavity in the ground state $|0\rangle$. The result of a spontaneous emission event during the cycle is therefore to map the initial state $|1\rangle$ to the final state $|0\rangle$ with probability $p$. The action of spontaneous emission on a state $\rho$ that is being transferred is therefore given by the operation 
\begin{equation}
   \rho = |\psi\rangle \langle \psi |  \rightarrow   E_0 \rho E_0 +  E_1 \rho E_1 
\end{equation}
in which 
\begin{eqnarray}
  E_0 & = & \sqrt{1 - p}\,\sigma_+\sigma_-, \\ 
  E_1 & = & \sqrt{p}\,\sigma_- ,   
\end{eqnarray}
where $\sigma_- \equiv |0\rangle \langle 1|$ and $\sigma_+ \equiv |1\rangle \langle 0|$. Denoting this operation by $\mathcal{E}$, which we think of as a noisy communication channel (also referred to as an amplitude damping channel), we can write the coherent information as  
\begin{equation}
   \mathcal{I}(\rho,\mathcal{E}) = S(\mathcal{E}[\rho]) - \mathcal{S}(\rho,\mathcal{E}) , 
\end{equation}
in which $S$ is the von Neumann entropy, 
\begin{equation}
   \mathcal{E}[\rho] =  \sum_{n=0}^1 E_n \rho E_n, 
\end{equation}
and $\mathcal{S}(\rho,\mathcal{E})$ is called the \textit{entropy exchange}. This last quantity measures the entropy that is fed into the environment (and thus lost) by the channel. The entropy exchange is defined by 
\begin{equation}
   \mathcal{S}(\rho,\mathcal{E}) = - \mbox{Tr}[W \ln W]   
\end{equation}
in which the elements of the matrix $W$ are given by 
\begin{equation}
  W_{jk} = \mbox{Tr}[E_j \rho E_k] .    
\end{equation}

Writing the density matrix $\rho$ as 
\begin{equation}
   \rho = \left(
\begin{array}{cc}
1-q & c\\
c^* & q \\
\end{array}\right) ,
\end{equation}
we find that the coherent information for the decay channel $\mathcal{E}$ is given by 
\begin{equation} 
   \mathcal{I}(\rho,\mathcal{E}) = \frac{1}{2}\log \left[ \frac{(1 - B)^{1-B}(1+B)^{1+B}}{(1-A)^{1-A}(1+A)^{1+A}} \right], 
\end{equation} 
in which 
\begin{align}
   A & =  \sqrt{\left(2 (p-1) q +1\right){}^2-4 (p-1) |c|^2} \\
   B & =  \sqrt{\left(1-2 p q\right){}^2+4 p |c|^2} .  
\end{align} 
If we use a base two logarithm then we obtain the coherent information in units of qubits. 

\begin{figure}[tb] 
\includegraphics[width=\columnwidth]{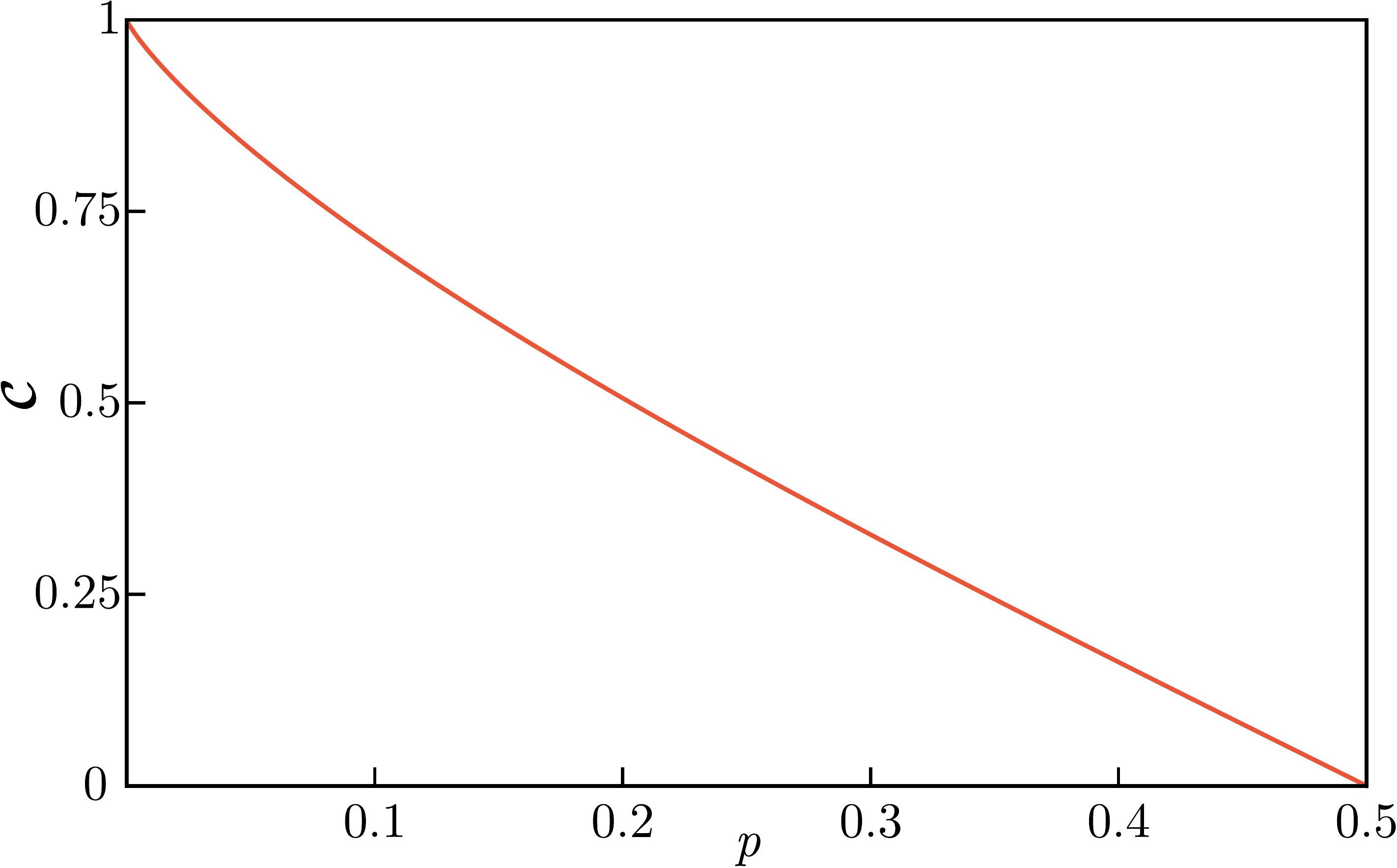}
\caption{(Color online) The quantum communication capacity of the transfer of a single photon, measured in qubits, as a function of the spontaneous emission probability, $p$.}
\label{fig:cap1}
\end{figure}

To determine the quantum information capacity of the channel as a function of the decay probability $p$, which we will denote by $\mathcal{C}(p)$, we must maximize the coherent information over all initial states $\rho$ for each value of $p$. We perform this optimization numerically and plot the channel capacity as a function of $p$ in Fig.~\ref{fig:cap1}. The \textit{rate} at which quantum information can be transmitted by the frequency converter depends on the time it takes to convert a single excitation. This conversion time depends, in turn, on the control protocol we use to mediate the transfer cycle. For a given conversion time, $T$, we can find the protocol that achieves the minimal loss probability, $p$, and write this as $p(T)$. In this way we parameterize the control protocols using the conversion time. We can then write the rate of reliable quantum communication for a given conversion time as  
\begin{align}
   \mathcal{R}[p(T)] =  \frac{1}{T} \mathcal{C}[p(T)] =  \frac{1}{T} \max_{\rho} \mathcal{I}[\rho,\mathcal{E}(p[T])] . 
\end{align} 
The fastest communication rate is then given by maximizing $\mathcal{R}[p(T)]$ to find the optimal conversion time $T$. 

\section{Physical system}
\label{secPS}

\begin{figure*}
\includegraphics[width=1.8\columnwidth]{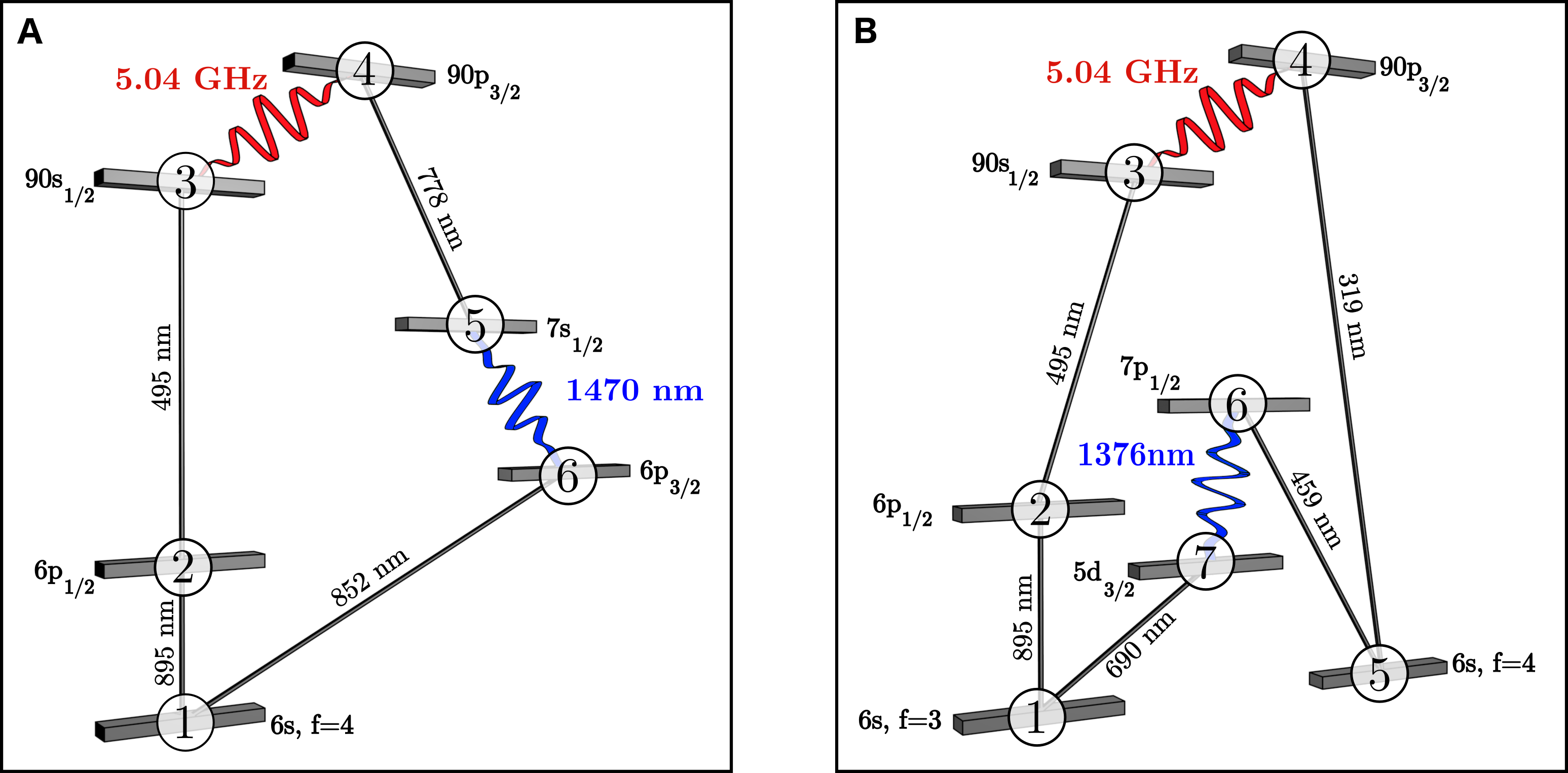}
\caption{(Color online) Here we show two configurations in which sets of levels of the cesium atom are coupled in a closed loop (or ``cycle'') by laser fields, a single mode of a superconducting resonator, and a single mode of an optical cavity.   Cycle A contains six atomic levels in which one is a ground state, while cycle B contains seven levels with two grounds states. The lifetimes of the levels, wavelengths of the transitions, and the maximum coupling rates between the levels are given in Tables~\ref{tbl:params1} and~\ref{tbl:params2}. The Rydberg transitions that couple to the microwave resonator are denoted in red, and the optical transitions are denoted in blue. For each level we show the orbital designation as well as a simple numbering scheme that we use in our analysis.}
\label{fig:atoms} 
\end{figure*}

Our system, depicted in Fig.~\ref{system_pic}, consists of a microwave waveguide resonator~\cite{cpw1}, a trapped cesium atom, and an optical cavity. One atomic transition couples to a microwave mode and one to an optical mode. To achieve frequency conversion we utilize lasers to couple the microwave transition to the optical transition via intermediate levels to form a closed loop; see Fig.~\ref{fig:atoms}. If a microwave photon is absorbed by the microwave transition (level $3\rightarrow4$) a laser coupled transition will take the atom from a Rydberg level of the microwave transition (level $4$) to the top level of the optical transition (level $5$), allowing it to emit a photon into the optical cavity and eventually return to the level at which it started (level $1$). In Fig.~\ref{fig:atoms}(a-b) we show two configurations (cycles) in which this loop structure can be achieved. 

The primary limitations on the transmission rate are given by the respective coupling rates between the atom and the two resonators, and by the maximum values of the coupling rates (the Rabi frequencies) that can be obtained with the lasers. Trapping the cesium atom at a distance of $\sim 10~\mu\rm m$ from the surface, and using a microwave resonator with frequency $\omega_{\ms{m}} =2 \pi \times 5.04~\mbox{GHz}$, provides a single-photon coupling rate of $g_{\ms{m}} = 2 \pi \times 3~\mbox{MHz}$~\cite{cpw1}. Placing the atom in an optical cavity with dimensions $L=70~\mu\rm m$ and using $g=d \sqrt{\omega/(2\epsilon_0\hbar V)}$ with $\omega=2\pi c/\lambda$ the angular frequency, $d$ the reduced radial matrix element, and $V= w^2 L $ the mode volume where $w=\sqrt{\lambda L/2\pi}$ is the cavity confocal waist, we find a single-photon coupling rate of $g_{\ms{o}}=(2 \pi \times 200~\mbox{MHz}, 2 \pi \times 100~\mbox{MHz})$ for cycles A and B, respectively. Assuming realistic laser powers not exceeding 10 mW, the maximum Rabi frequencies that can be attained for the various transitions are shown in Tables~\ref{tbl:params1} and~\ref{tbl:params2}, and span the range $20\text{--}1200~\mbox{MHz}$. The two main factors that limit the speed of the conversion (the speed at which the loop can be traversed) are the smallest coupling strength, which in our case is the coupling to the microwave resonator, and the need to minimize the spontaneous emission from the atomic levels in the loop. To achieve the latter, as we will explain below, one must use the laser fields to realize adiabatic transfer processes that are much slower than the maximum Rabi frequencies. We find that it is the first of these that sets the speed of communication.  

The Hamiltonian that describes the resonator-atom-resonator system, in the interaction picture with respect to the Hamiltonians of the individual systems, is given by 
\begin{align} 
   H = \,\, & \hbar \sum_{jk \in \mathcal{C}} \Omega_{jk}(t) \exp \left[-i \Delta_{jk}(t) t \right] \sigma_{jk} + \mbox{H.c.} \nonumber \\ 
         & + \hbar \left[ g_{\ms{m}} a^\dagger \sigma_{\ms{m}}  +  \hbar g_{\ms{o}}  b^\dagger \sigma_{\ms{o}} \right]  +  \mbox{H.c.}   
\end{align} 
Here the raising and lowering operators for the atomic transitions that are coupled by the lasers are defined by $\sigma_{jk} = |j\rangle \langle k|$. The transitions that couple to the microwave and optical modes are denoted by $\sigma_{\ms{m}}$ and $\sigma_{\ms{o}}$, respectively. (Thus for cycle A $\sigma_{\ms{m}} = |3\rangle \langle 4|$, $\sigma_{\ms{o}} = |6\rangle \langle 5|$ and for B $\sigma_{\ms{m}} = |3\rangle \langle 4|$, $\sigma_{\ms{o}} = |7\rangle \langle 6|$.) Here $\mathcal{C}$ is the set of pairs $(j,k)$ that represent the pairs of levels coupled by the lasers, and we are working in the interaction picture with respect to the Hamiltonians of the individual systems. Thus for cycle A these pairs are $\{ (1,2), (2,3), (4,5), (1,6) \}$ and for B we have $\{ (1,2), (2,3), (4,5), (5,6), (1,7) \}$. The frequencies $\Delta_{jk}$ give the amounts by which the lasers are detuned from their respective transitions. The annihilation operators for the microwave and optical cavities are denoted by $a$ and $b$, respectively.

Apart from the ground states, each atomic level $|j\rangle$ decays with rate $\gamma_j$. If a decay occurs during the conversion cycle the excitation being transferred is lost and the atom must be pumped back into the appropriate ground state to begin the cycle again. Under the appropriate weak-damping (rotating-wave) approximation the evolution of the joint system is described by the master equation~\cite{scully97, wang2012} 
\begin{align} 
   \dot\rho = \,\, &  - \frac{i}{\hbar}[H,\rho] + \sum_{j\in \mathcal{D}} \gamma_j L(\sigma_{0j}) [\rho] +\kappa L(b)[\rho]
\end{align} 
in which $\mathcal{D}$ is the set of levels that decay, $\kappa$ is the decay rate of the optical cavity, and the action of the Lindblad super-operator $L(c)$ on $\rho$ is defined by 
\begin{align} 
   L(c) [\rho] =   c \rho c^\dagger - \frac{1}{2}\left( c^\dagger c \rho + \rho c^\dagger c \right) . 
\end{align} 
In the above master equation we have defined the new lowering operators $\sigma_{0j} \equiv |0\rangle \langle j|$. Each of these describes a transition from level $|j\rangle$ to a new state $|0\rangle$ that we have introduced merely to catch the population that is lost in the transfer process. 

Our central problem is to determine how to choose the constant coupling rates $g_{\ms{m}}$ and $g_{\ms{o}}$, the Rabi frequencies $\Omega_{jk}(t)$, and the detunings $\Delta_{jk}(t)$ over a time interval $[0,\tau]$ so as to complete one of the cycles depicted in Fig.~\ref{fig:atoms} with the smallest loss probability, $p$. Note that by definition $p$ is a property of the transfer process (the operation $\mathcal{E}$ in Section~\ref{secQTR}) rather than of the state being transferred. As such we can determine $p$ by starting with a single photon in the microwave cavity and the vacuum in the optical cavity, and calculating the probability that the optical cavity contains a photon at time $\tau$. (Note that the actual probability of a decay occurring during the transfer depends both on $p$ and the initial state in the microwave mode.) 

In Tables~\ref{tbl:params1} and~\ref{tbl:params2} we give the lifetimes of the atomic levels, the wavelengths of the transitions, and the constraints that we impose on the size of the Rabi frequencies (or the coupling rates to the cavity modes, whichever is appropriate), for cycles A and B, respectively. 

\begin{table}[t]
\begin{tabular}{|c|c|c|c|c|}
\hline 
Level & $1/\gamma$ (ns) & $\lambda$ or $f$  &P (mW) &  Max. Coupling Rate\\
\hline
1  & $\infty$ & 895  nm& $0.001$ &$2 \pi \times630$ MHz   \\
2 &34.8 & 495  nm& 10 &$2 \pi \times70$ MHz\\
3 & $8\times 10^5$ & $5.04~\rm  GHz$   & - &$2 \pi \times3$ MHz   \\ 
4 & $2\times 10^6$  &  778 nm &10 &$2 \pi \times 50$ MHz  \\
5 & $48$ & 1470 nm  &- &$2 \pi \times 200$ MHz  \\
6 & $30.4$ & 852 nm &0.001 &$2 \pi \times  880$ MHz  \\
\hline 
\end{tabular}
\caption{Parameters for the cesium levels and transitions in cycle A. For level $n$ the wavelength $\lambda$ and Rabi frequency constraint $\Omega$ are for the transition from level $n$ to the next level in the cycle. The Rabi frequency constraint was calculated using the reduced radial matrix element between the levels, neglecting angular factors, and  assuming the indicated optical power focused to a spot with Gaussian waist ($1/e^2$ intensity radius) of $2~\mu\rm m$. For the 1470 nm transition the coupling rate is the vacuum Rabi frequency calculated from $g=d \sqrt{\omega/(2\epsilon_0\hbar V)}$ with $\omega=2\pi c/\lambda$ the angular frequency, $d$ the reduced radial 
matrix element, and $V= w^2 L $ the mode volume where $L=70~\mu\rm m$ is the cavity length and $w=\sqrt{\lambda L/2\pi}$ is the cavity confocal waist. For the 5.04 GHz transition $g$ was calculated using the parameters in \cite{quantumgates2}.} 
\label{tbl:params1}
\end{table} 

\section{Physics of the conversion process}
\label{secTPO}

Ideally one would like to determine the fastest reliable rate of quantum communication given the constraints of the system. The speed of the transfer of a single photon is the speed at which the cycle of atomic levels can be traversed, and this is limited by the coupling between the levels (given by the coupling to the cavities and the lasers). Insight is provided by noting that there are two other effects that limit the conversion speed, and over which one must optimize to maximize this speed. One of these is due to the decay rates of the cavities, and the second is the need to minimize the probability of loss via spontaneous emission. 

\subsection{Cavity damping rates and conversion time}

The microwave and optical cavities provide the interface between the atom and the systems that will use the quantum information being converted. Once the atomic cycle has successfully transferred a photon from one cavity to the other, we must have a mechanism to extract the photon (and thus the quantum state) from either cavity. The conversion time, $T$, is given by 
\begin{equation}
   T = \tau + t_{\ms{io}} , 
\end{equation}
in which $t_{\ms{io}}$ is the time required to load the photon into, and output it from, the cavities. 

For the microwave cavity it is entirely reasonable to envisage a superconducting qubit that can be effectively coupled and decoupled from the cavity so as to insert and extract the quantum state when needed. Further, this time can be short compared to the transfer time and so we ignore it. For the optical cavity, on the other hand, the simplest configuration is to have the cavity damp directly to an optical fiber that may, for example, transmit the photon over a long distance. In this case the cavity is permanently coupled to its output. In view of this we need to take two effects into consideration. The first is that the time taken to output the photon into the transmission line is random, and we must wait until we are fairly sure this output has occurred before initiating another transfer. We therefore want to choose a cavity lifetime that is significantly shorter than the transfer time $\tau$. The second effect is that the optical cavity damping rate, $\kappa$, modifies the dynamics of the transfer cycle. Interestingly, and from previous analyses~\cite{Law97, Cui05, Jacobs08, Wang11}, it is clear that when the damping rate is larger than the coupling rate between the atom and the cavity, the measurement aspect of the damping induces a quantum Zeno effect that inhibits the transfer. This slows the transfer process and in doing so increases the probability that the photon will be lost to spontaneous emission. However, as we have already discussed, if the damping rate of the optical cavity is smaller than the transfer rate, then the photon will take a significant time to exit the cavity and reduce the overall transmission rate. We can expect, therefore, that there is an optimal choice for the damping rate of the optical cavity. 

\begin{table}[t]
\begin{tabular}{|c|c|c|c|c|}
\hline
Level & $1/\gamma$ (ns) & $\lambda$ or $f$ &P (mW) &  Max. Coupling Rate\\
\hline
1  & $\infty$ & 895  nm& $0.001$ & $2\pi \times630$ MHz  \\
2 &34.8 & 495  nm& 10 & $2\pi \times70$ MHz  \\
3 & $8\times 10^5$ & $5.04~\rm  GHz$   & - & $2\pi \times 3$ MHz   \\
4 & $2\times 10^6$  & 319  nm & 10 & $2\pi \times  20$ MHz \\ 
5 & $\infty$ & 459 nm& 1  &$ 2\pi \times 1200$ MHz\\
6 & $155$ & 1376  nm& -&$ 2\pi \times 100$ MHz \\
7 & $910$ & 690  nm& 10 &$ 2\pi \times 250$ MHz\\
\hline
\end{tabular} 
\caption{Parameters for the cesium levels and transitions in cycle B. The definitions are the same as in Table 1. For the quadrupole transition from level 7 to 1 we calculate the Rabi frequency using the reduced matrix element of $r^2$ times a factor of $2\pi a_0/\lambda$ with $a_0$ the Bohr radius. For the 1376 nm transition the coupling rate is the vacuum Rabi frequency calculated from $g=d \sqrt{\omega/(2\epsilon_0\hbar V)}$ with $\omega=2\pi c/\lambda$ the angular frequency, $d$ the reduced radial 
matrix element, and $V= w^2 L $ the mode volume where $L=50~\mu\rm m$ is the cavity length and $w=\sqrt{\lambda L/2\pi}$ is the cavity confocal waist.}
\label{tbl:params2}
\end{table} 

To find optimal transfer protocols under the constraints on the laser power we must fix the time allowed for the protocol, as described below, and thus determine an optimal protocol separately for each transfer time. Because the cavity damping rate contributes separately to the total time required for communication (by setting the time taken to transfer a photon from the cavity to the transmission line), optimizing this rate would also require separate optimizations for each value of the damping. An exhaustive optimization over both transfer time and damping rate is thus rather prohibitive. 

To handle this problem we perform an optimization for a comprehensive range of protocol durations $\tau$ but do this for just two values of the cavity decay rate. First we choose a cavity decay rate that we expect to provide near-optimal transfer efficiency. (Since the resulting transfer efficiency and communication rate can in theory be achieved, these automatically provide a lower bound on the achievable values.) To select a near-ideal value for the cavity decay rate we note first that the rate of the transfer will be limited by the smallest of the coupling rates that make up the cycle. In our case this is the coupling to the microwave cavity which is much smaller than that to the optical cavity. This means that there is no need to make the optical cavity damping rate much larger than the optical coupling rate, since at this point the time to exit the cavity will already be smaller than the transfer time. Finally, we do not expect the damping of the optical cavity to significantly slow the transfer unless the damping rate is at least similar in size to the coupling to the optical cavity. These three facts together indicate that, so long as $g_{\ms{o}}/g_{\ms{m}}$ is sufficiently large, the optimal value for $\kappa$ will satisfy $g_{\ms{m}} < \kappa < g_{\ms{o}}$. 

In view of the fact that we must wait long enough between transfers to be sure that the photon has exited the optical cavity (and thus a number of cavity lifetimes), it is desirable to have $\kappa$ on the high end of the range we have identified. We note also that Cui and Raymer~\cite{Cui05} have shown that when driving a two-level atom coupled to a cavity with coupling rate $g$, the choice of $\kappa$ that maximizes the probability that photons are emitted from the cavity (rather than from the atom via spontaneous emission) is $\kappa = 2 g$. While the above arguments indicate that this value is a little on the high side for our situation, we perform our optimizations with 
\begin{equation}
   \kappa = 2 g_{\ms{o}} . 
\end{equation}
Since the damping rates of microwave cavities can be made much smaller than their optical counterparts, and in view of the fact that photons can be extracted from microwave cavities by a variety of means, we set the damping rate of the microwave cavity to zero. 

The value we selected above for $\kappa$ will provide us with a transfer efficiency (success probability) that can, in theory, be achieved. Nevertheless, we don't know that this is optimal so it is also worth placing an upper bound on the achievable efficiency. Since we expect the cavity decay to only inhibit the transfer, setting $\kappa=0$ can be expected to provide such an upper bound, so we also perform an optimization for this case. 

\subsection{The transfer process and STIRAP}

Once we have fixed the cavity damping rates and chosen a transfer time, the task of the numerical optimization is to tailor the shapes of the laser pulses to minimize the probability that the photon is lost from the atom during the transfer. In fact, there is already a well-known method that enables high-fidelity population transfer between two stable levels when the population must pass through a lossy intermediate level. This method is known as ``stimulated Raman adiabatic passage'' (or STIRAP for short)~\cite{stirap}. In our analysis here, while we let the numerical search find what protocols it will, it turns out that the properties of the resulting optimal protocols indicate that the primary mechanism they exploit is that of STIRAP, so we review it briefly now. 

Lets say that we wish to transfer population from state $|\mbox{A}\rangle$ to $|\mbox{B}\rangle$ via a third lossy state $|\mbox{L}\rangle$. A STIRAP process would function by adiabatically transferring population by way of a dark state, which minimally populates the lossy state $|L\rangle$. Since STIRAP is an adiabatic process the transfer is only perfect in the limit of a long transfer time. The energy gap between the decoupled dark state that carries the population and the other states is set by the maximum size of the Rabi frequencies, and thus by the maximum available laser power. The total loss thus increases with the transfer speed, and for a fixed value of the loss the speed increases with the available laser power. While the transfer process we need for frequency conversion involves more than three levels, it is evident that the optimal protocols exploit the adiabatic STIRAP mechanism: for a given maximum Rabi frequency the faster we perform the transfer the greater the resulting loss. 

\begin{figure*}[!tb]
\includegraphics[width=2.06\columnwidth]{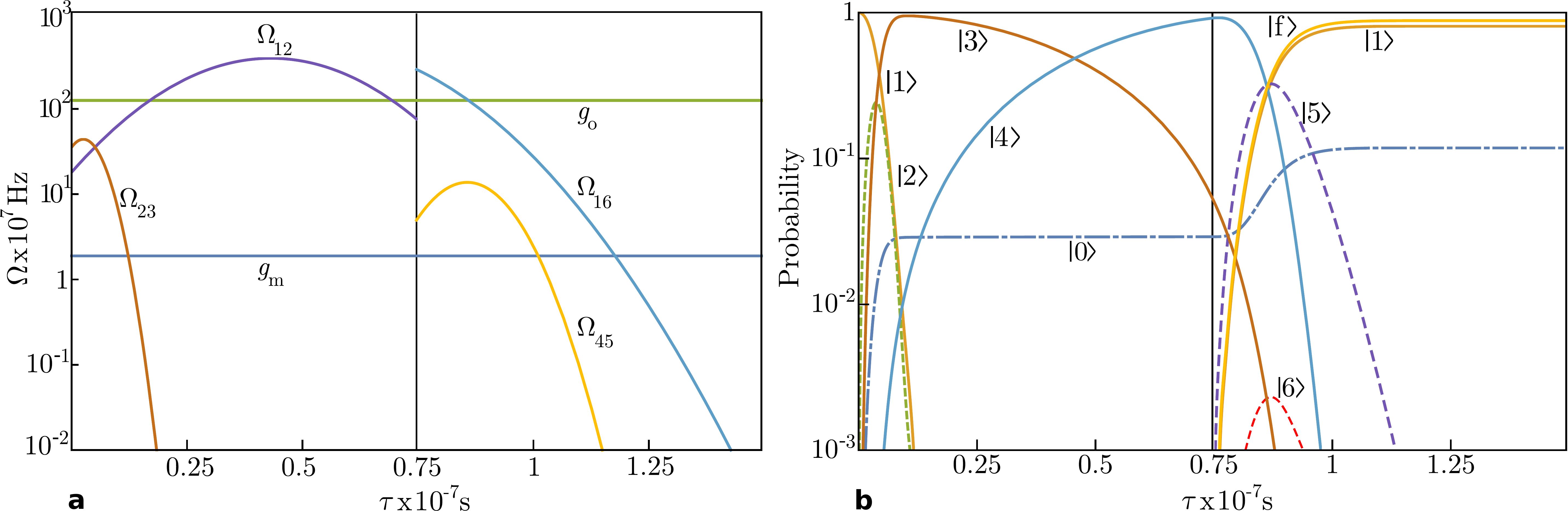}
\caption{(Color online) Here we show an example of a control protocol found by our numerical search for cycle A, along with the evolution of the atomic populations that it generates in traversing the cycle once. (a) The time-dependent envelopes for the Rabi frequencies and the atom/mode coupling rates that constitute the control protocol. Note that the protocol is divided into two halves. Any Rabi frequencies not shown on the plot are zero during that half. (b) The populations of the atomic levels over the conversion cycle. Note that the levels with the fastest decoherence rates are quickly traversed. Level f represents the sum of successfully reaching the optical cavity and the subsequent travel out of the cavity.}
\label{fig:gaussian}
\end{figure*}

\section{Numerical optimization}
\label{nopt} 

To find optimal transfer protocols we fix the duration, $\tau$, allowed for the transfer, and the decay rate of the optical cavity, and perform a numerical optimization to search for time-envelopes for the Rabi frequencies $\Omega_{jk}$ and detunings $\Delta_{jk}$ that provide the minimal loss probability, $p$. We then repeat this procedure for a range of durations to find $p$ as a function of $\tau$. To perform numerical optimization we must write these envelopes in terms of a finite set of real parameters to obtain a finite-dimensional search space. We want to both minimize the numerical overhead required for the simulation, as well as obtain some confidence that our protocols are near-optimal while at the same time minimizing the size of the search space. We now describe the approach we have taken to these problems. 

\textit{Efficient simulation.---}The full system consists of two oscillators, each with two accessible states, and an atom with either 6 or 7 states. Fortunately we can significantly reduce the size of the space required for the simulation by noting that if one resonator starts with one photon and the other starts in the vacuum state not all of the joint states of the combined systems are accessible during the evolution. Further, since any spontaneous emission from the atomic levels represents a failure of the transfer, we can correctly account for this failure by having every atomic level decay independently to a single auxiliary level (that subsequently stores the failure probability). Of course we must also include the fact that the photon, once transferred to the optical cavity, can decay to the output. Once again, including this decay requires only one additional auxiliary level. Note that at the start of the transfer process the atom is in the ground state and the microwave and optical modes are in the 1-photon and vacuum states, respectively. Once the atom returns to the ground state (in the absence of spontaneous emission) the states of the modes have been reversed, and thus the initial and final states of the ``cycle'' are distinct. By enumerating all the joint states that are accessible from the initial state, and including the two auxiliary levels, we find that with $n$ atomic levels in the cycle the simulation requires only $n+4$ states. The total probability that the transfer succeeds is the population of the 1-photon state of the optical cavity at the end of the protocol, plus the population that has decayed from the cavity into the transmission line (and is thus stored in the second auxiliary level).  

It turns out that we can, in fact, reduce the complexity of the optimization by taking advantage of the fact that the Rydberg levels have very long lifetimes. This means that after the atom has absorbed the microwave photon we can leave the population in the upper Rydberg level for some time without any significant effect on the success of the transfer. This allows us to break the optimization problem into two segments. In the first segment we optimize the transfer from the initial ground state to the upper Rydberg level. In the second segment we optimize the rest of the transfer. We find that breaking the optimization into two segments reduces the resulting success probability by only about 1\%. 

\begin{figure*}[!t]
\includegraphics[width=2.06\columnwidth]{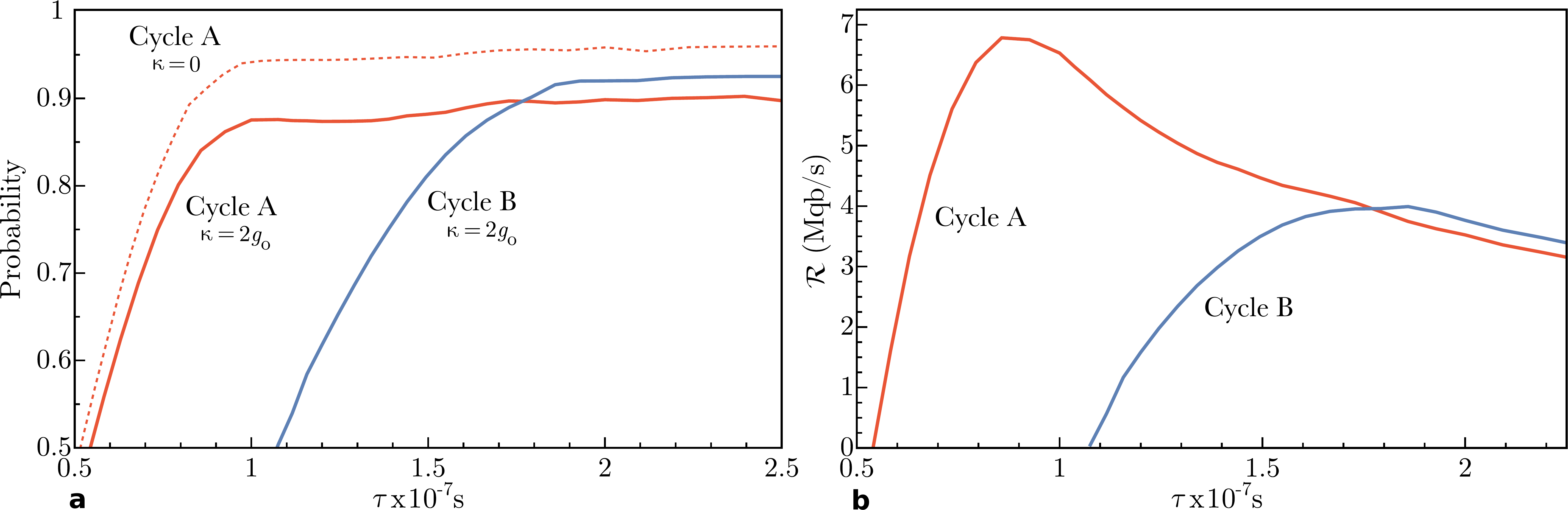}
\caption{(Color online) (a) Probability of the successful conversion of a single photon as a function of the cycle duration (the single-photon transfer time) $\tau$, and as obtained by our numerical optimization (using Gaussian-parametrized envelopes as discussed in the text). For the solid curves the optical cavity decay rate is $\kappa=2g_o$ while the dashed curve shows the performance for $\kappa=0$ . (b) The quantum communication rate (capacity per unit time) that results from the transfer probabilities shown in (a). We see that there is an optimum value of the single-photon transfer time that maximizes this rate. The maximum communication rate is $\mathcal{R}=6.814~\mbox{Mb/s}$ at $\tau=85.7~$ns for cycle A and $\mathcal{R}=4.009~\mbox{Mb/s}$ at $\tau=185.9~$ns for B. Red: cycle A; Blue: cycle B.}
\label{fig:results1}
\end{figure*}

\textit{Parameterizations and optimization.---}We begin by setting all the detunings to zero and optimizing the Rabi frequencies and the coefficients $g_{\ms{m}}$ and $g_{\ms{o}}$ that couple to the modes. We use first a piecewise constant parametrization of the envelopes. Specifically, we divide the interval $[0,\tau]$ into $N$ segments of respective durations $\tau_j$ (with $\sum_j \tau_j = \tau$). Each envelope is constant within each interval, but envelope $k$ is allowed to take a different value $v_{jk}$ for each interval. We then optimize over the parameters $\tau_j$ and $v_{jk}$. We find that performance is poor for $N=1$, increases dramatically when going from $N=1$ to $N=2$, and then increases only a little when increasing $N$ to 3. This indicates that two intervals is sufficient to obtain the majority of the performance. In contrast to previous works, which typically assume all Rabi frequencies are constant in time \cite{jaksch17, jaksch16}, we find that using time dependent pulses can greatly improve conversion efficiency. As a double-check and to employ a parametrization that is less discontinuous, we choose each envelope to be a Gaussian with a variable mean, amplitude, and width. This provides quite a flexible functional form, allowing a peak, an upward curve, a downward curve, or something close to flat. We found that optimization using Gaussian envelopes gives very similar results to those obtained using piecewise constant envelopes with $N=2$. Finally, for various values of $\tau$, we performed optimizations using piecewise constant pulses with $N=2$ in which we allowed not only the Rabi frequencies to vary but also the detunings within a range $\pm 10 \mbox{MHz}$. Our results indicate that no benefit is obtained by detuning the lasers from their respective transitions. 

While there are a large number of numerical search algorithms we could employ, we find that a Nelder-Mead simplex search performs well in terms of computation time and accuracy~\cite{opt99}. Naturally, any optimization algorithm attempts to find a set of optimal parameters but finds any non-unique set that satisfies the algorithm constraints. Performing our search multiple times, we then gain confidence that our search returns a set which is likely optimal.

\section{Results}
\label{secR}

In Fig.~\ref{fig:gaussian} we display the pulse envelopes and the resulting population dynamics for a single example of an optimized transfer using cycle A. Note that the protocol is divided into two parts as described above. During the first half of the protocol the Rabi frequencies $\Omega_{16}$ and $\Omega_{45}$ are turned off (and thus the coupling rate $g_{\ms{o}}$ is irrelevant) so as to transfer from level $|1\rangle$ to $|4\rangle$, and in the second half the Rabi frequencies $\Omega_{12}$ and $\Omega_{23}$ are turned off. In each half of the protocol the pulse envelopes are optimized under the restriction that each has a Gaussian form. We see from the plots that the populations of the most rapidly decaying states ($|2\rangle$, $|5\rangle$, $|6\rangle$) are occupied only a little, and for relatively short periods during the cycle, so that the most stable levels carry the majority of the population. We note also that the envelopes of the Rabi frequencies in the first half of the protocol have the form of STIRAP envelopes, in that the coupling for the second transition ($\Omega_{23}$) peaks before that of the first ($\Omega_{12}$). 

We also examine the advantage provided by the use of time-dependent, optimized pulse shapes over merely driving the cycle transitions with constant laser power. To do so we simulate the evolution with constant driving, this time performing an optimization over the choice of driving power for each laser. As an example, choosing a protocol time of $\tau=150~\mbox{ns}$ we find that while the optimized Gaussian pulses achieve a transfer probability of $P\approx87.8\%$, constant driving manages only $P\approx37.7\%$. 

We now perform the optimization for both cycles A and B over a range of durations $\tau$ to determine the duration that achieves the maximum probability of a successful transfer (maximum efficiency). We present the results of these optimizations in Fig.~\ref{fig:results1}(a). The solid lines in Fig.~\ref{fig:results1}(a) show the efficiencies achieved when $\kappa = 2 g_{\ms{o}}$. We see that while both A and B achieve similar maximum efficiency, cycle A achieves this at significantly shorter durations. Thus the resulting transmission rate, shown in Fig.~\ref{fig:results1}(b) is significantly higher for A than for B. As we have noted above, the choice $\kappa = 2 g_{\ms{o}}$ is not necessarily optimal. In Fig.~\ref{fig:results1}(a) we also show the success probability achieved by cycle A when the decay rate of the optical cavity is zero. As discussed above, this is expected to provide an upper bound on the achievable performance. 

Given the transfer probability (as a function of the transfer time) we can calculate the quantity of primary interest, the quantum communication capacity per unit time (being the rate at which the conversion process can be used to transmit quantum information reliably). To do so we must also decide how long to give the optical cavity to output the converted photon to the transmission line between each transfer. Choosing this time to be $\tau_{\ms{io}} = 10/\kappa$, the resulting probability that the cavity fails to output each converted photon is less than $10^{-4}$. In Fig.~\ref{fig:results1} we display the resulting quantum transmission rates for cycles A and B as a function of the protocol transfer time, $\tau$. We see that as we allow the protocol more time to perform the transfer a point is reached at which the cost of the increased time outweighs that of the increased success probability, and the maximum communication rate is reached. The maximum communication rates are 
\begin{eqnarray}
   \mbox{Cycle A: } & & \;\; \mathcal{R} = 6.78~\mbox{Mqb/s} \;\;  \mbox{at} \; \tau=85.7~\mbox{ns} , \nonumber \\ 
   \mbox{Cycle B: } & & \;\; \mathcal{R} = 3.99~\mbox{Mqb/s} \;\;  \mbox{at} \; \tau=185.9~\mbox{ns} , \nonumber
\end{eqnarray}
in which Mqb/s denotes Mega-qubits per second. 

\begin{figure}[t] 
\includegraphics[width=\columnwidth]{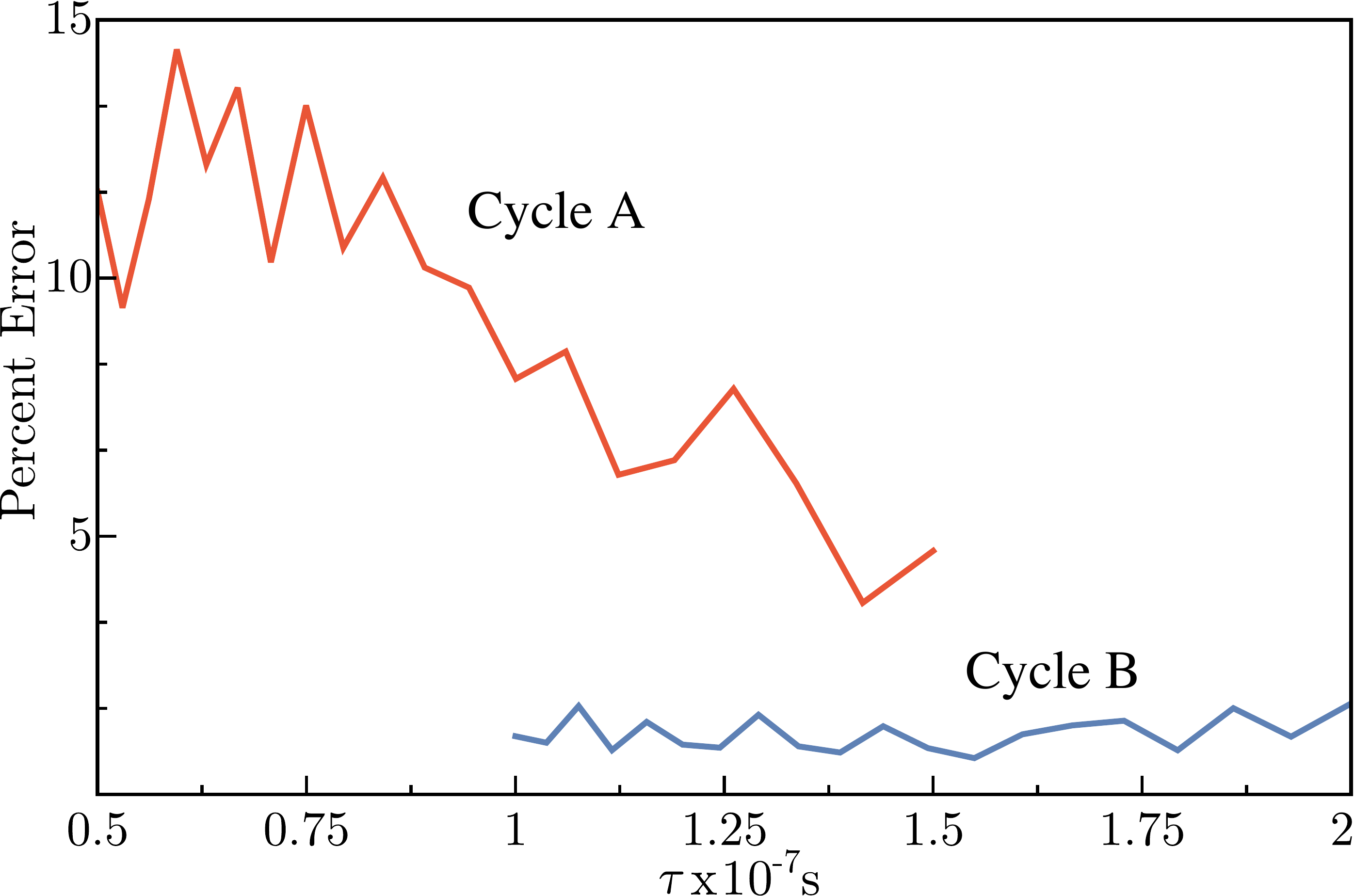}
\caption{Here we show the (average, fractional) increase in the loss probability when each parameter in the control protocol is subject to Gaussian noise. The settings of this Gaussian noise are fixed to have mean and standard deviation $(\mu,\sigma_\mu=\mu/25)$, where $\mu$ is each optimal parameter setting. The averages are taken over two hundred values of the error and over all the control parameters. These results are for the protocols that use the Gaussian-parameterized envelopes as discussed in the text. We have truncated both cases as each cycle configuration has a significantly different relevant timescale, as is clear from Fig.~\ref{fig:results1}. Red: cycle A; Blue: cycle B.} 
\label{fig:error}
\end{figure}

To complete our analysis we check the \textit{robustness} of the control protocols that we have obtained for the cyclic transfer process. A highly optimized protocol has no practical use if its performance is dramatically degraded by imperfections in the envelopes of the control parameters. To confirm that small imperfections cause only small reductions in performance we simulate the protocols with noise added to the envelopes. Specifically we add to each envelope a constant offset chosen from a Gaussian distribution whose mean and standard deviation are set by $(\mu,\sigma_\mu=\mu/25)$, where $\mu$ is each optimal parameter setting. We perform simulations for two hundred samples of this noise applied separately to each Rabi frequency, and average the resulting deviations in the transfer probability over these simulations. In Fig.~\ref{fig:error} we plot the average deviation in the transfer probability as a function of the transfer time, confirming that moderate errors in the protocols cause similarly moderate errors in the performance. 

Throughout our analysis we have mainly considered our conversion device as converting from microwave to optical, rather than the other way around. Naturally it will be important to convert from optical to microwave as well. We have performed numerical optimizations of the transfer process in this direction, in which case we reverse the order of our two-part optimization so that the optical photon is extracted before being placed in the microwave mode. We find that the resulting transfer times are very similar, and so we have not included separate plots for them. However, there is an important caveat when considering optical to microwave conversion. Note that in treating the reverse process (microwave-to-optical) we can assume that the lifetime of the microwave cavity is long compared to the transfer time. This means that we can assume the state stays safely in the microwave mode while the transfer is completed. As we have discussed in some detail, due to technological limitations we assume that the optical cavity is permanently damped to the optical transmission line with a damping rate that is much faster than the transfer time. This is good for microwave-to-optical conversion but not for the reverse direction. Further, if damping rate of the optical cavity is fixed (cannot be changed with time) then one must take into account that the photon could leak back out of the optical mode during the transfer and minimize this possibility. To do this one would need to include the optical transmission line in the treatment (which could be done, for example, by including the system that sends the photon over this line) and the transfer protocol would be optimized both to capture the photon from the line (see, e.g.~\cite{Cirac97}) as well as perform the conversion. In this case we can expect that the result will be lower final communication rates than those we have obtained here. An analysis of the dynamics of the transmission line and/or the sending system in the conversion process is beyond our scope here, but may be a worthwhile subject for future work. 

\section{Conclusion}
We have undertaken an analysis of a quantum frequency converter that uses a single cesium atom as its core element. We have shown that by optimizing the laser pulses used to couple the atomic levels to perform the conversion efficiencies of over 90\% are theoretically possible with realistic laser powers and other design parameters. The resulting rate for reliable quantum communication is on the order of a few Mega-qubits per second. In our treatment we have also allowed some asymmetry between the microwave mode and the optical mode as regards technological difficulties. In particular we have assumed that the lifetime of the microwave mode can be chosen to be long without restricting the input and output from this mode, while the lifetime of the optical mode is set by the need to use this damping channel to send the photon to the optical output line. We have found that the transfer process itself, as performed by having the atom traverse a cycle of levels, provides similar performance in both directions. We have explicitly calculated the quantum communication rate assuming that the source mode has a long lifetime, which with present technology is appropriate for the microwave cavity, and thus for microwave-to-optical conversion. If the optical cavity is restricted to a damping rate that is similar to the atom transfer rate (for example to facilitate input/output to the optical transmission line), then conversion rates in the other direction will be lower. We have pointed out that in this case a full optimization of the performance for optical-to-microwave conversion would require inclusion of the optical transmission line and/or the sending system. 

\begin{acknowledgments}
This research was supported in part by an appointment to the Postgraduate Research Participation Program at the U.S. Army Research Laboratory administered by the Oak Ridge Institute for Science and Education through an interagency agreement between the U.S. Department of Energy and USARL. This research was also supported by Army Research Office Contract W911NF-16-1-0133 and by the US Army Research Laboratory Center for Distributed Quantum Information through cooperative agreement W911NF-15-2-0061. 
\end{acknowledgments}

%
% Bibliography
%
%\bibliography{bib,report_QMT}


\end{document}